\documentclass[]{aa} % for the long lists of affiliations 
\usepackage[colorlinks=true, linkcolor=blue, citecolor=blue, urlcolor=blue]{hyperref}
\usepackage{amsmath}
\usepackage{graphicx}
\usepackage{amssymb}
\usepackage{mathrsfs}
\usepackage{caption}
\usepackage{float}
\usepackage{afterpage}
\usepackage{amstext}
\usepackage{multirow}

\usepackage{txfonts}
\usepackage{natbib}
\defcitealias{Pakstiene2026}{Paper~I}
\begin{document}
%
%}
   \title{The non-LTE abundances of magnesium and yttrium and asteroseismic ages for the chemical clock calibration
   %\thanks{Based on }
}
   \author{
\v{S}. Mikolaitis\inst{1},  
G. Tautvai\v{s}ien\.{e}\inst{1}, 
E. Pak\v{s}tien\.{e}\inst{1},
A. Drazdauskas\inst{1},
V. Bagdonas\inst{1}, \\
C. Viscasillas V\'azquez\inst{1},
M. Ambrosch\inst{1},
Y. Chorniy\inst{1},
R. Minkevi\v{c}i\={u}t\.{e}\inst{1}, 
E. Stonkut\.{e}\inst{1}, \\
B. Bale\inst{1},
B. Ćurjurić\inst{1},
A. Sharma\inst{1},
K. Diktanait\.{e}\inst{1}
}

   \institute{Vilnius University, Faculty of Physics, Institute of Theoretical Physics and Astronomy, Saul\.{e}tekio av. 3, LT-10257 Vilnius, Lithuania\\
              \email{sarunas.mikolaitis@tfai.vu.lt } %1
             }

   \date{Received 2026/ Accepted 2026}

\authorrunning{\v{S}. Mikolaitis et al.}
\titlerunning {The NLTE abundances of magnesium and yttrium and asteroseismic ages for the chemical clock calibration}

% \abstract{}{}{}{}{} 
% 5 {} token are mandatory
 
  \abstract
   {}
  % aims heading (mandatory)}
{Building on our previous study, which demonstrated the necessity of accounting for departures from the local thermodynamic equilibrium (LTE) while determining elemental abundances and the importance of asteroseismic ages, our aim with this work was to characterise the spatial variations of the empirical [Y/Mg]--age relation across the Galactic disc with a significantly larger sample of stars.  
}
{We observed high-resolution stellar spectra and determined Mg and Y abundances via spectral synthesis of multiple spectral features and rigorously accounted for non-LTE (NLTE) effects. To anchor the timescale, we determined asteroseismic ages for stars showing solar-type pulsations while employing cross-checked isochrone-based methods for the remaining stars.}
   {We determined the main atmospheric parameters and abundances of Mg and Y for a sample of 528 Galactic field stars as well as asteroseismic ages for 307 stars and cross-checked isochronal ages for 221 stars. We identified two new triple-lined and nine double-lined spectroscopic stellar systems. Based on a total sample of 736 stars, with data from this study and our previous one, we explored the [Y/Mg] versus age relations across the Galactic disc. }
  {The [Y/Mg] versus age relations exhibit systematic variations across the Galactic discs, reflecting differences in star formation and enrichment histories. There is a tendency for [Y/Mg] to increase with increasing metallicity across the age range. However, at supersolar metallicity this tendency may not hold, and the relations become flatter compared to solar metallicity stars, which have lower [Y/Mg] values at young ages and higher values at old ages.}


   \maketitle
%________________________________________________________________

\section{Introduction}
Chemical clocks have become a widely studied phenomenon. The ability to somehow derive or at least infer ages of stellar objects from their chemical abundances is a promising  method. Overall, the chemical-clock method reveals completely new possibilities of investigation since age is one of the most important parameters in astronomy. However, age is very difficult to determine precisely. Developing this new chemical-clock method would allow astronomers to not only derive ages of various stars but also investigate ages of much larger objects, such as the thin and thick disc of the Galaxy, as well as the timescales of various processes.

During the past decade, the [Y/Mg] ratio has been established as a very promising chemical clock. There are two main reasons for this. First, the origins of these elements are completely different. Magnesium is synthesised predominantly in massive stars and released into the interstellar medium on short timescales via core-collapse supernovae. As Type~Ia supernovae began to contribute significantly to the chemical enrichment of the interstellar medium by injecting large quantities of iron-peak elements, the [Mg/Fe] ratio steadily declined over the course of Galactic evolution. Yttrium, in contrast, is synthesised predominantly through the $s$-process in asymptotic giant branch stars on longer timescales, leading to a gradual increase in [Y/Fe] towards younger stellar populations. Second, while there are many elements produced in the $s$-process, yttrium is among the elements with the largest fraction of the $s$-process in its synthesis \citep{Arlandini99, Simmerer04, Sneden08, Bisterzo14,Prantzos20}. Moreover, its abundances in various studies are much more confined and robustly determined. 
For example, barium has the largest $s$-process fraction, and the [Ba/Mg] ratio could therefore be a more pronounced chemical clock \citep{2021A&A...652A..25C, Vazquez22}, but barium shows a phenomenon known as the barium puzzle and is difficult to analyse \citep{Reddy17}. Cerium, another element with a  high $s$-process fraction in its production, has also been studied as a potential chemical clock in combination with $\alpha$-elements \citep[e.g.][]{Casali23}. However, up until now, [Y/Mg] has attracted the most attention.

Among the first studies to define [Y/Mg] as a potential chemical clock, along with other ratios, was the work by \citet{daSilva12}. The study showed that for solar-type stars, [Y/Mg] increases significantly towards younger objects. A similar conclusion has been reached through the study of solar twins (\citealt{Nissen15, 2016A&A...593A.125S, 2016A&A...590A..32T, 2018MNRAS.474.2580S, Jofre20}), where the precision of age determination may reach \textasciitilde0.8~Gyr or even \textasciitilde0.5~Gyr. Subsequent studies concluded that the [Y/Mg] ratio can be used as a chemical clock for solar metallicity stars (\citealt{Nissen16, Adibekyan16}). Using asteroseismic age estimates for a sample of 66 solar-type $Kepler$ LEGACY stars observed with high-resolution spectroscopy, \citet{Nissen2017} demonstrated that the [Y/Mg] abundance ratio exhibits a tight correlation with stellar age. Later, the same tendency was found in core-helium burning stars \citep{2017A&A...604L...8S}. 

The [Y/Mg] relation with age was expanded to FGK dwarfs in a study by \citet{Delgado19}, where more than 1000 objects were used to establish the [Y/Mg] correlation with age in different metallicity ranges. A study of FGK dwarfs in the solar neighbourhood by \citet{Feltzing17} showed that as metallicity decreases, the [Y/Mg] relation with age flattens. In the study of the turn-off thin-disc stars by  \citet{2019A&A...622A..59T}, the dependence of [Y/Mg] versus age relation on metallicity was not found. 
\citet{Vitali24} used 74 field giants with asteroseismic ages and determined that there are different [Y/Mg] trends for stars of different metallicity. \citet{Berger2022} investigated FGK-type planet-host stars and confirmed that [Y/Mg] as an age indicator is most precise for solar twins and analogues but stated that it can be applied for FGK stars as well. 

\citet{Casali20} studied open clusters from the $Gaia$-ESO survey and came to the conclusion that the [Y/Mg] ratio is not universal throughout the Galaxy. This was also confirmed by \citet{2021A&A...652A..25C} based on the investigation of red clump stars in open clusters. 
\citet{Magrini2021} adopted a new set of asymptotic giant branch stellar yields with magnetic mixing and also came to the conclusion that [Y/Mg] cannot be universal as a stellar clock throughout the Galactic disc. 
 \cite{Vazquez22, Vazquez25}, and \cite{Ratcliffe24} found that the [Y/Mg]–age relation varies with the Galactocentric distance, becoming flatter in the inner regions and steeper towards the outer disc. 

The study by \citet{Tautvaisiene2021} expanded the investigation of FGK stars by including the thick-disc representatives and concluded that the [Y/Mg] versus age relation works for the thin but not for the thick disc.  \citet{2024A&A...690A.107S} employed a sample of solar-type stars, among them solar twins and solar analogues, and confirmed that [Y/Mg] depends on metallicity. However, they stated that the thin and thick discs have a similar [Y/Mg] correlation with age. 

Although quite a number of studies have investigated [Y/Mg] as a chemical clock, many questions remain. 
It is therefore important to investigate further chemical clocks across the Galactic disc or at different metallicity regimes. Furthermore, Y, especially for more metal poor giant stars \citep{Storm2023}, and to a lesser extent Mg \citep{Bergemann2017} need non-local thermodynamic equilibrium (NLTE) effects to be taken into account. Finally, the robustness of the [Y/Mg] versus age relation can be improved by using the asteroseismic ages for calibration.  

Recently, \cite{Pakstiene2026}, hereafter called \citetalias{Pakstiene2026}, conducted a study of 208 field stars by determining the asteroseismic stellar ages and [Y/Mg] ratios with NLTE effects taken into account. It was demonstrated that the [Y/Mg] versus age relation exhibits a clear radial dependence across the Galactic disc, with a steeper trend in the outer disc, progressively flatter relations towards the inner disc, and a very flat trend in the thick disc. It was emphasised that the NLTE abundances of Mg and especially of Y must be used to have a more precise stellar age evaluation from [Y/Mg] ratios.

In this work, we extend the sample of stars investigated in \citetalias{Pakstiene2026} with high-resolution spectral observations of 528 field stars and their Mg and Y abundances derived with NLTE effects taken into account. In 307 stars, solar-type pulsations were strong enough to determine ages asteroseismically. For other stars, the ages were also determined by using the isochrone-based method. With this expanded sample of 736 stars, we address spatial differences in the [Y/Mg]-age relation.

\section{Observations and method of analysis}
\label{sec:3}

\subsection{Observations}

We observed 626 spectra of 563 F-, G-, and K-type stars with $V < 8$~mag that were monitored by the Transiting Exoplanet Survey Satellite (TESS) satellite, extending the efforts of \citet{Tautvaisiene2020,Tautvaisiene2022} to observe bright stars in the northern hemisphere.
The observations were carried out with the Vilnius University Echelle Spectrograph (VUES; \citealt{2014SPIE.9147E..7FJ,Jurgenson16}) mounted on the 1.65~m Ritchey–Chrétien telescope at the Molėtai Astronomical Observatory, Lithuania. VUES covers the wavelength range 4000–9000~\AA. Two resolving powers, $R$, of $\sim 68\,000$ and $\sim 36\,000$ were used for the observations, with the higher one used for most of the targets. Signal-to-noise ratios, measured per pixel, range from 35 to 298, with the vast majority of spectra having values between 60 and 110.

Spectral reductions were performed using standard spectroscopic techniques.
Bias, flat-field, and calibration-lamp measurements were acquired every evening before observing the stellar spectra. We used a quartz lamp for flat-fielding and ThAr spectra for wavelength calibration. Data were reduced and calibrated following standard procedures, including bias subtraction, flat-field correction, extraction of echelle orders, wavelength calibration, and cosmic-ray removal \citep{Jurgenson16}.

\begin{figure}
    \centering
    \includegraphics[width=1.0\columnwidth]{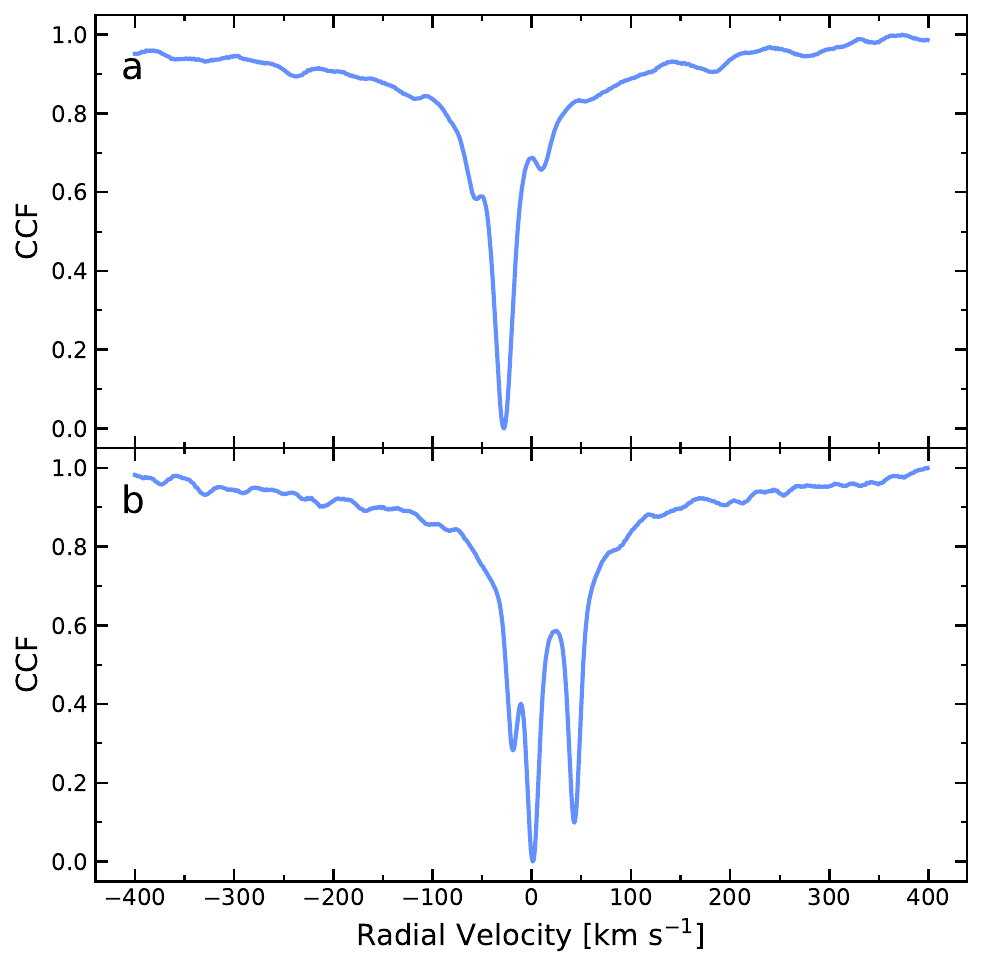}
    \caption{Example CCFs showing triple-line features in the spectra of stars (a) TYC~3893-469-1 and (b) TYC~2995-190-1.}
    \label{fig-CCF-example}
\end{figure}

Since the target selection was based only on photometric indices, the initial sample was expected to include a number of fast rotators and spectroscopic double-line binaries. We therefore performed a cross-correlation of each spectrum with a numerical mask constructed from an atomic line list in order to identify multi-lined spectroscopic systems and determine radial velocities. The mask was built using the Gaia-ESO Survey line list compiled by the Gaia-ESO line list group \citep{Heiter2021}. For each star, the cross-correlation function (CCF) was computed over the radial-velocity range $-400$ to $+500~\mathrm{km\,s^{-1}}$ using the steps of $\Delta V_{\mathrm{rad}} = 0.1~\mathrm{km\,s^{-1}}$. 

During the spectral reduction, we identified 20 stars with projected rotational velocities $v \sin i > 20$~km\,s$^{-1}$; nine new double-lined spectroscopic binaries (TYC2551-233-1, TYC3082-784-1, TYC3083-618-1,  TYC3318-1840-1,  TYC3550-914-1, TYC3691-1248-1, TYC3885-130-1, and TYC4492-1319-1, TYC5105-837-1) along with three previously known binaries (TYC1268-6-1, TYC1954-1881-1, and TYC3138-1424-1, TYC3358-3141-1); and two new triple-lined spectroscopic systems (TYC2995-190-1 and TYC3893-469-1; see Fig.~\ref{fig-CCF-example} for an example). These stars were excluded from further processing. Thus, the final analysis was performed for 528 remaining stars. 

\subsection {Stellar atmospheric parameters and chemical element abundances} 
\label{sec:chemistry}

Stellar atmospheric parameters ($T_{\rm eff}$, log\,$g$, [Fe/H], and $V_{\rm t}$) were derived from equivalent widths of up to 86 Fe\,{\sc i} and 7 Fe\,{\sc ii} lines using the same classical spectroscopic techniques as in \citetalias{Pakstiene2026}. Spectral lines were selected from an initial set of 299 Fe lines in the Gaia-ESO Survey atomic data compilation of \citet{Heiter2021} including all lines with synflag = Y and lines with synflag = U that also have gfflag = Y. Additional selection criteria related to line quality, blending, and line strength were applied to obtain the final line list (Table~\ref{table:AtomicData}).

Effective temperatures were obtained by minimising trends of Fe\,{\sc i} abundances with the excitation potential, surface gravities from ionisation equilibrium, and microturbulent velocities by requiring the Fe\,{\sc i} abundances to be independent of the equivalent widths. The analysis was performed with the 19$^{\rm th}$ version of the MOOG code \citep{1973PhDT.......180S}, MARCS LTE stellar atmosphere models \citep{Gustafsson08}, and the solar composition by \cite{2007SSRv..130..105G}. The median uncertainties are $\sigma T_{\rm eff}=\pm47$~K, $\sigma{\rm log}g=\pm0.16$, $\sigma{\rm [Fe/H]}=\pm0.09$, and $\sigma v_{\rm t}=\pm0.19$~km\,s$^{-1}$.

Magnesium and yttrium abundances were determined by spectral synthesis of the lines presented in Table~\ref{table:AtomicData} using the Turbospectrum (v. 2019)  code  \citep{1998A&A...330.1109A,Plez2012} with the solar composition by \cite{2007SSRv..130..105G} in a line-by-line differential analysis relative to the Sun observed with the same spectrograph and telescope. For stars observed at two spectral resolutions, solar spectra with matching resolutions were used. For the resolution $\sim$36\,000, the averaged abundances of Mg and Y for the Sun were $7.56\pm 0.12$~dex and $2.19\pm 0.12$~dex, respectively; for the resolution $\sim$68\,000, these values were $7.54\pm 0.09$~dex and $2.18\pm 0.13$~dex, respectively. The values are very close to the solar abundances of Mg ($7.53\pm 0.09$~dex) and Y ($2.21\pm 0.02$~dex) presented by \citep{2007SSRv..130..105G}. For the investigated stars, abundances were averaged over multiple observations when available. Individual uncertainties are listed in Table~\ref{table:Results} along with the resulting stellar Mg and Y abundances.

The total abundance uncertainty of the Mg and Y abundances for each star was calculated as
\\
$e_{\rm ([X/H])} = \sqrt{\sigma^2_{\rm total([X/H])}+\left(\frac{\sigma_{\rm N}}{\sqrt{N}}\right)^2   } $,
\\
where $\sigma_{\rm total([X/H])}$ represents the uncertainty resulting from the adopted atmospheric parameters, $\sigma_{\rm N}$ is the line-to-line scatter, and $N$ is the number of spectral lines analysed. For magnesium, the median line-to-line scatter is 0.06~dex, while the median abundance uncertainty due to atmospheric parameter uncertainties is 0.07~dex. Combining these contributions yielded a median total abundance uncertainty of 0.09~dex. For yttrium, the median line-to-line scatter is 0.07~dex, and the median abundance uncertainty caused by atmospheric parameter uncertainties is 0.08~dex, resulting in a median total abundance uncertainty of 0.11~dex.

For the abundances of magnesium and yttrium, we computed the NLTE corrections according to \cite{Storm2023} and \cite{Storm2024} for Y\,{\sc ii} and according to \cite{Bergemann2017} for Mg\,{\sc i} using the updated Turbospectrum code (\citealt{Gerber2023}).
 The NLTE effects on the [Mg/H] abundances in the  investigated metallicity interval are small. However, on [Y/H], they reach $\sim 0.08$~dex at the lowest metallicities. 

\subsection{Orbital characterisation}
\label{sec:distances and discs}

\begin{figure}
    \centering
    \includegraphics[width=0.49\columnwidth]{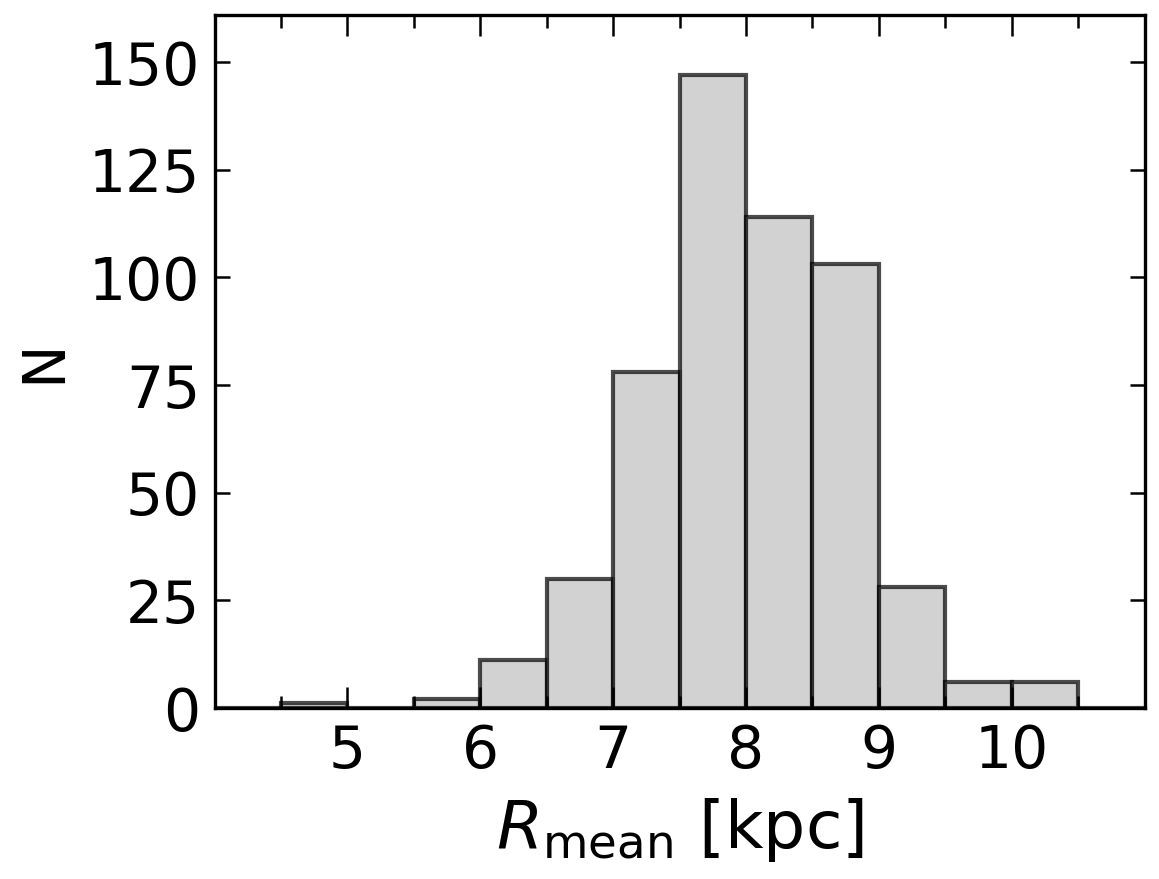}
    \includegraphics[width=0.49\columnwidth]{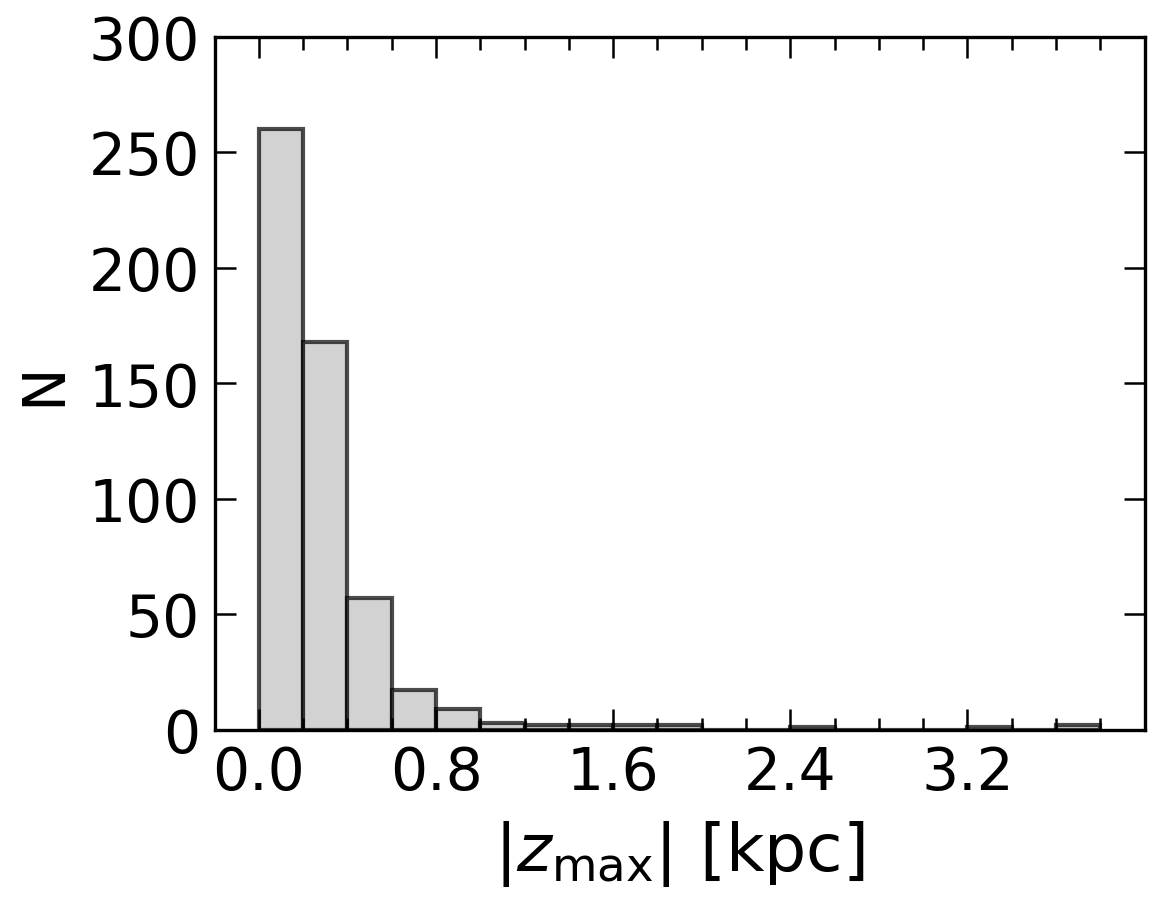}
    \caption{Distributions of the stellar orbital parameters of the full sample of investigated stars. }
    \label{histo-kinem}
\end{figure}

To characterise the dynamical properties of our sample, we derived the mean Galactocentric distance ($R_{\text{mean}}$) and the maximum excursion from the Galactic plane ($|z_{\text{max}}|$) as well as the velocities $U$, $V$, $W$ following the methodology outlined in \citetalias{Pakstiene2026}. Orbital integrations were performed using the Python-based galactic dynamics library \texttt{galpy} \citep{Bovy15}. 

\begin{figure}[]
    \centering
    \includegraphics[width=9cm]{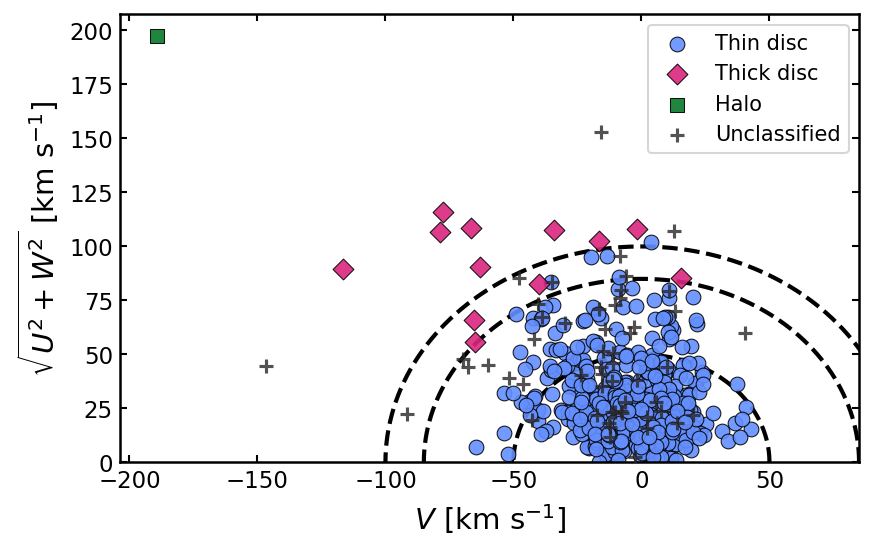}
    \caption{Toomre diagram of the investigated stars, attributed to the Galactic discs. The blue circles are for the thin disc, the magenta diamonds are for the thick disc, the green square is for the halo star, and the grey crosses are for the unclassified stars.}
    \label{fig:Toomre}
\end{figure}

\begin{figure}[]
    \centering
    \includegraphics[width=9cm]{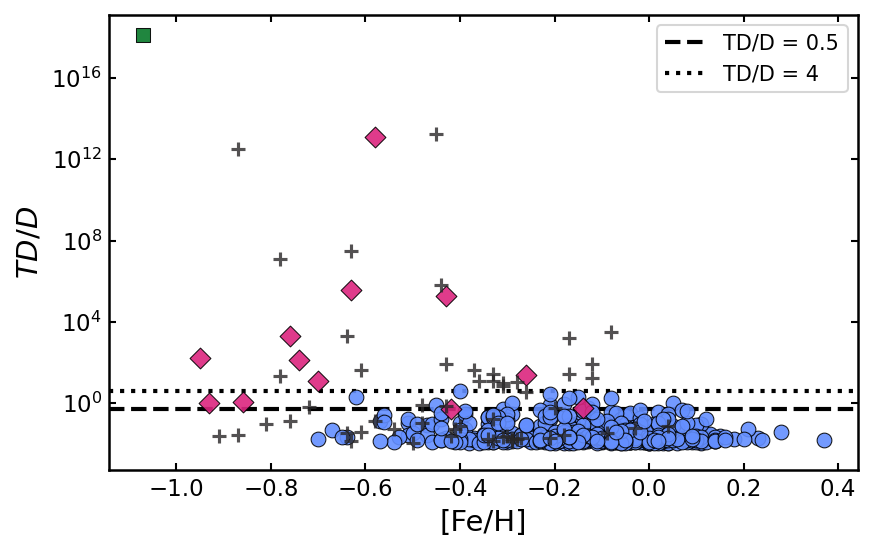}
    \caption{Kinematical thick-to-thin disc probability ratio ($TD/D$) versus metallicity. The two horizontal lines denote the boundaries at $TD/D$~$<$~0.5 and $TD/D$~$>$~4 used for the separation of the thick- and thin-disc stars. Symbols are as in Fig.~\ref{fig:Toomre}.}
    \label{fig:TDD}
\end{figure}

To perform \texttt{galpy} calculations, we needed to know the object position and velocities in 3D space. The initial conditions for the orbits were constructed using right ascension, declination, and proper motions retrieved from the {\it Gaia} DR3 catalogue (\citealt{GaiaDR3}). The distances were taken from \cite{bailerjones21}, and the radial velocities were determined directly from the spectra. We adopted the axisymmetric \texttt{MWPotential2014} galactic potential provided within \texttt{galpy}. The orbit integrations were performed over 5~Gyr using 10\,000 time steps, corresponding to a time resolution of 0.5~Myr, to ensure stable estimates of orbital parameters. For these calculations we adopted a default solar velocity value of 220~km\,s$^{-1}$ and the galactocentric distance of $R_\odot = 8.0$~kpc from \cite{Bovy12} and the default height above the Galactic plane of $Z_\odot = 28$~pc \citep{bennett19}. By propagating observational astrometric and radial velocity errors using a Monte Carlo approach, we estimated a typical mean uncertainty of 0.04~kpc for $R_{\text{mean}}$ and 0.007~kpc for $|z_{\text{max}}|$. The distribution of $R_{\text{mean}}$ and $|z_{\text{max}}|$ for our sample stars is presented in Fig.~\ref{histo-kinem}.

To dynamically separate the sample into thin- and thick-disc populations, we calculated the space velocity components ($U, V, W$) relative to the local standard of rest (LSR). We adopted the peculiar solar motion of $(U, V, W)_{\odot}~=~(11.1, 12.24, 7.25)$~km\,s$^{-1}$ from \cite{schonrich10}. Furthermore, we calculated the relative likelihood of thick-to-thin disc membership ($TD/D$) while assuming the velocity components follow Gaussian velocity ellipsoids and using asymmetric drifts from \cite{Bensby14} and standard characteristic velocity dispersions ($\sigma_U, \sigma_V, \sigma_W$) from \cite{vieira22} for each population.

\begin{figure}[]
    \centering
    \includegraphics[width=9cm]{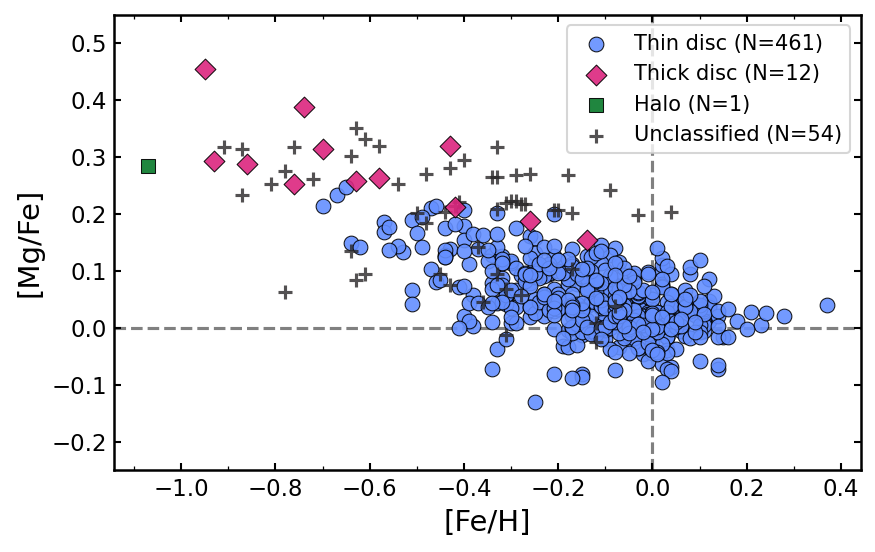}
    \caption{[Mg/Fe] versus [Fe/H] plot indicating stars of the thin and thick discs, the halo star, and the unclassified stars. Symbols are as in Fig.~\ref{fig:Toomre}.}
    \label{fig:MgFe}
\end{figure}

Because the Galactic thin and thick discs exhibit significant overlap in phase space, relying on a single separation criterion can introduce severe contamination. Therefore, we employed a strict threefold separation approach: (1) The combined space velocities had to exceed 85~km\,s$^{-1}$ (see Fig.~\ref{fig:Toomre}), (2) stars were required to have a kinematic probability ratio of $TD/D > 0.5$ (see Fig.~\ref{fig:TDD}), and (3) the star had to display an enhanced [Mg/Fe] signature separating them from the thin disc (see Fig.~\ref{fig:MgFe}).
Only stars that were classified as members of the Galactic thick disc according to all three criteria were attributed to the Galactic thick disc in our study. Three criteria for selecting thin-disc stars were applied as well: The combined space velocities had to be less than 100~km\,s$^{-1}$, stars were required to have a kinematic probability ratio $TD/D < 4$, and the star had to display non-enhanced [Mg/Fe]. In this way we divided the sample stars into 461 thin-disc and 12 thick-disc candidates. One star was attributed to the halo population, and 54 remained unclassified. 

\subsection{Stellar ages}

\subsubsection{Asteroseismic ages}

We analysed publicly available light curves from TESS (\citealt{Ricker15}). The photometric data were obtained from the SPOC pipeline \citep{2016SPIE.9913E..3EJ} products and from the TESS-SPOC High Level Science Products available through the Mikulski Archive for Space Telescopes (\citealt{MAST}).
The light curves were retrieved using the Python package \verb|lightkurve| \citep{LightkurveCollaboration2018} by querying the corresponding \verb|author| collections (`SPOC' and `TESS-SPOC') without imposing any cadence restrictions. Consequently, all available cadences for each target up to December 2025 were downloaded and analysed.

The SPOC light curves, available only for pre-selected targets, offer higher photometric precision and typically a 2-min cadence, making them generally preferable for asteroseismic analysis due to lower instrumental noise and more homogeneous systematic corrections, whereas the TESS-SPOC light curves, derived from full-frame images, cover a substantially larger number of targets with a wider range of cadences but generally lower precision and increased scatter.
For each target, we therefore selected the highest-quality light curve available, prioritising cadence and noise properties appropriate for the detection of solar-like oscillations.

The light curves were sigma-clipped using a combination of standard procedures available within the \verb|lightkurve| package. 
Light curves from all available sectors were combined prior to analysis to improve the signal-to-noise ratio and frequency resolution, which is particularly important for low-frequency oscillations in red giant stars.

The frequency at maximum oscillation power ($\nu_{\rm{max}}$) was estimated by fitting a Gaussian profile to the continuum-flattened power spectrum and identifying the frequency corresponding to the peak of the fit. The large frequency separation ($\Delta\nu$) was determined using the autocorrelation of the power spectrum in the frequency region around $\nu_{\rm{max}}$ and subsequently manually verified using échelle diagrams. The uncertainty in $\nu_{\rm{max}}$ was estimated from the full width at half maximum of the fitted Gaussian envelope of the oscillation power excess. The uncertainty in $\Delta\nu$ was estimated from the width of the autocorrelation peak and by visual inspection of the corresponding échelle diagram.

\begin{figure}
    \centering
            \includegraphics[width=0.49\columnwidth]{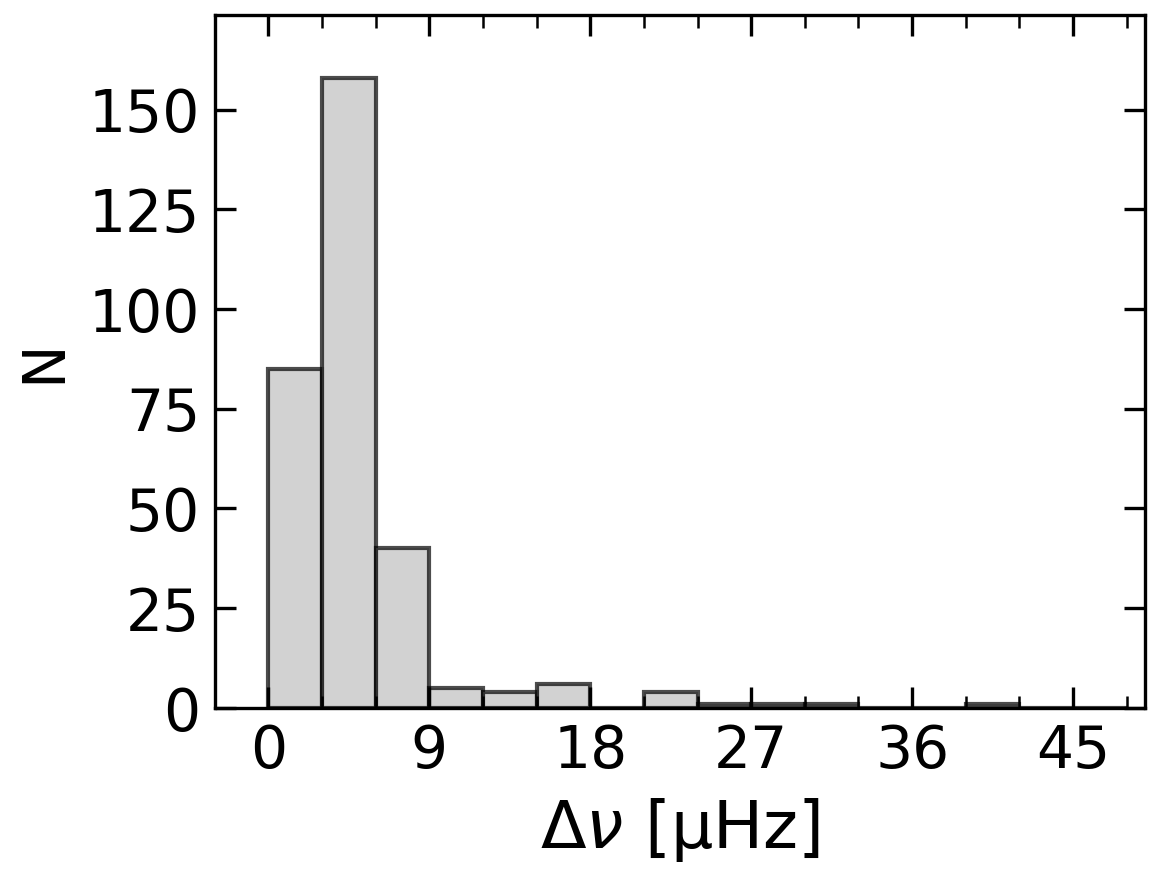}
    \includegraphics[width=0.49\columnwidth]{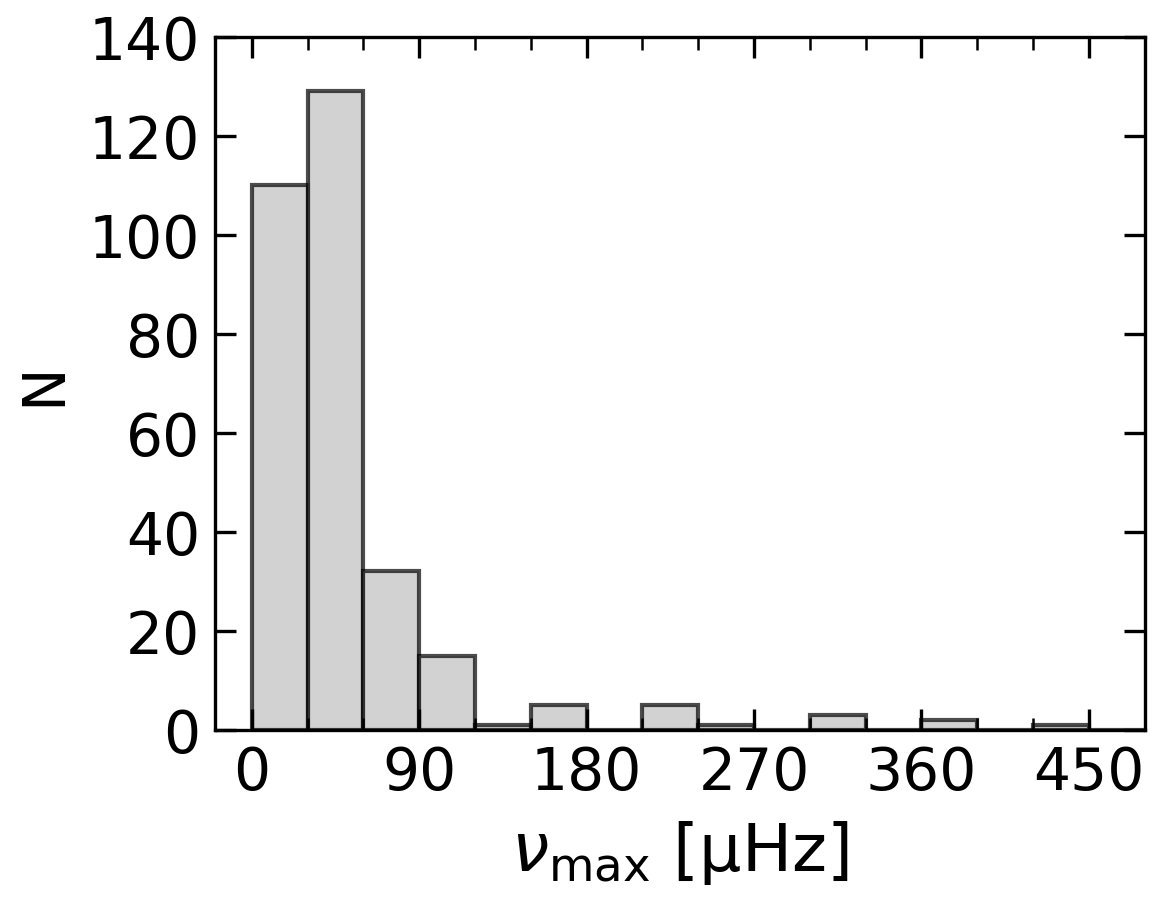}
    \caption{Distribution of $\nu_{\rm{max}}$ and $\Delta\nu$ determined.  }
    \label{histo-astero}
\end{figure}

We computed pulsation spectra for all stars, but robust determinations of $\nu_{\rm{max}}$ and $\Delta\nu$ were possible for just 307 stars. 
Only stars showing clear solar-like pulsations with easily recognisable ridges of $\ell =$ 0, 1, and 2 modes in échelle diagrams were selected for further analysis and age determination. Figure~\ref{histo-astero} 
shows the distribution of stars with $\nu_{\rm{max}}$ and $\Delta\nu$ determined. 

Using $\nu_{\rm max}$, $\Delta\nu$, and the spectroscopic effective temperature $T_{\rm{eff}}$, we calculated the initial approximate stellar mass, radius, $\log g$, and luminosity based on classical scaling relations \citep{KjeldsenBedding1995}.
All of these values along with spectroscopic $T_{\rm{eff}}$ and metallicity [Fe/H] were used as input parameters for the asteroseismic age calculation with an online interface PARAM (v.1.5) for the Bayesian estimation of stellar parameters \citep{daSilva2006, Rodrigues2014, Rodrigues2017}. We assumed the solar seismic parameters $\nu_{\rm max, \odot}=3141~\mu Hz$ and $\Delta\nu_\odot=134.98~\mu Hz$ \citep{FredslundAndersen2019}, a solar metal content of $Z_\odot=0.01756$ \citep{Paxton2011, Paxton2013}, an exponential initial mass function \citep{Chabrier2001}, and an unknown prior of the evolutionary stage. The solar metal content of $Z_\odot=0.01756$ was chosen because this value was implemented in MESSA isochrones used by the PARAM software. When using the $Z_\odot$ value recently provided by \citep{Lodders2025}, the ages would change by about 0.4\% for young stars ($\sim 1$~Gyr) and by about 1\% for older stars ($\sim 8$~Gyr).
To account for the mass loss in our calculations, we applied a Reimers-type mass-loss efficiency coefficient, $\eta_R = 0.4$. This value well reflects the average value of the Reimers coefficient considering different empirical contexts and models \citep{McDonald&Zijlstra2015, Valle&DellOmodarme2018}.

The asteroseismic ages were determined using a grid of MESA isochrones \citep{Rodrigues2017}, where individual radial mode frequencies and large separations $\Delta\nu$ were calculated for each model of the grid. As the PARAM description states that MESA isochrones are available in the mass range $0.6 \leq M/M_\odot < 2.5$ 
for stars with a scaling relation mass greater than $2.5 M_\odot$ (up to $4 M_\odot$), the age was calculated with the PARSEC isochrones \citep{Bressan2012} instead. Finally, a two-step Bayesian method \citep{Rodrigues2014} was used to determine the ages of our selected stars.

\subsubsection{Isochronal ages}

The isochronal ages were determined in this work with the Stellar Parameters INferred Systematically code (SPInS; \citealt{Lebreton2020}). SPInS is a Bayesian tool that compares observational constraints with grids of stellar evolutionary tracks and derives posterior probability distributions for age, mass, radius, and other parameters. It uses a Markov chain Monte Carlo sampler based on the \texttt{emcee} package \citep{ForemanMackey2013} to interpolate within the model grid.
As input constraints, we used our spectroscopic parameters $T_{\rm eff}$, $\log g$, and [Fe/H] together with photometric quantities: the absolute $V$-band magnitude ($M_V$) and the $(B-V)$ colour index. The $B$ and $V$ fluxes were compiled via the SIMBAD database from literature photometric catalogues \citep{Wenger2000}. We computed $M_V$ using Gaia~DR3 inverse parallaxes as a distance estimate and corrected the $B$ and $V$ magnitudes for reddening using three-dimensional dust maps of \cite{Green2019}. All constraints were supplied to SPInS with their respective uncertainties.
The uncertainties in the SPInS output ages for our sample correlate with the absolute age (Pearson coefficient $r \approx 0.74$), with an average uncertainty of 1~Gyr. Relative uncertainties do not show a correlation with age, and the mean relative age uncertainty for our sample is $\sim$30\%.

\section{Results and discussion}
\label{results}

\begin{figure}
    \centering
    \includegraphics[width=0.49\columnwidth]{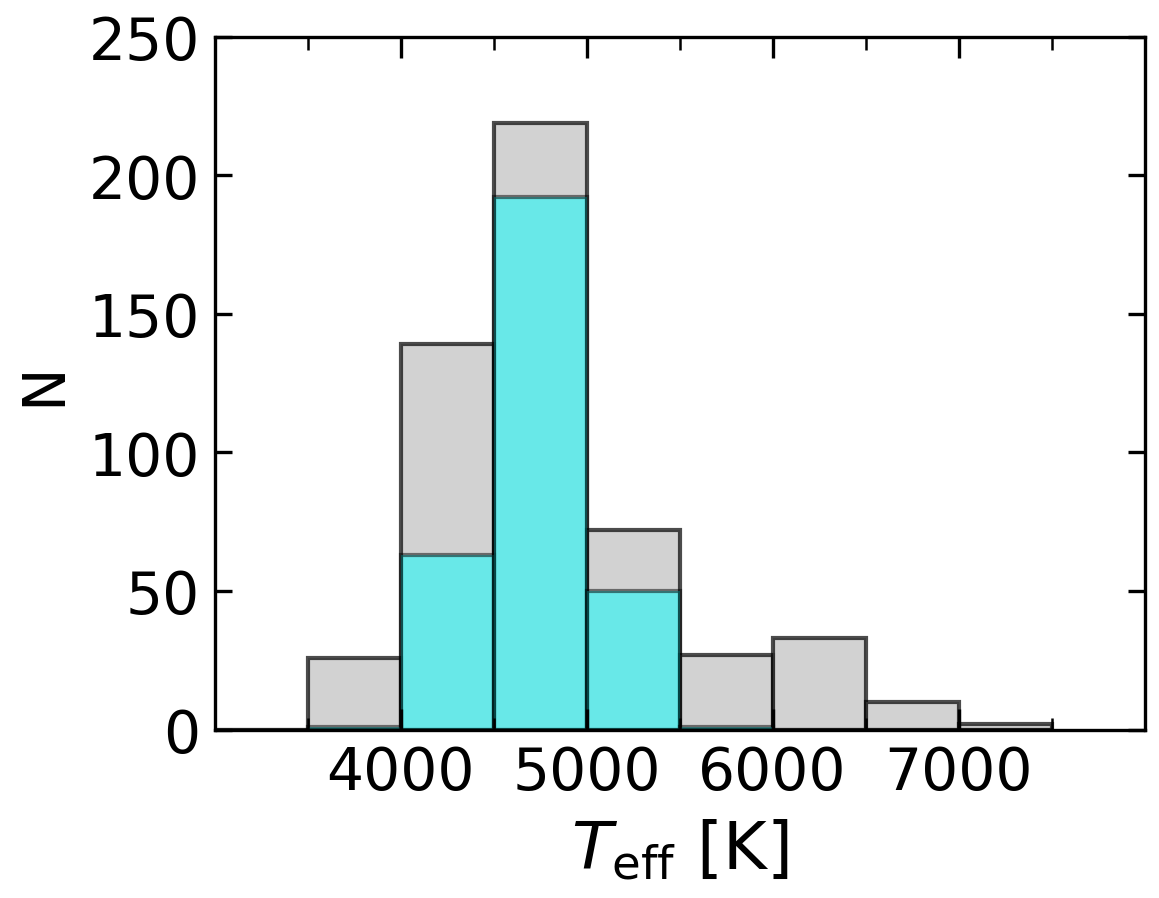}
    \includegraphics[width=0.49\columnwidth]{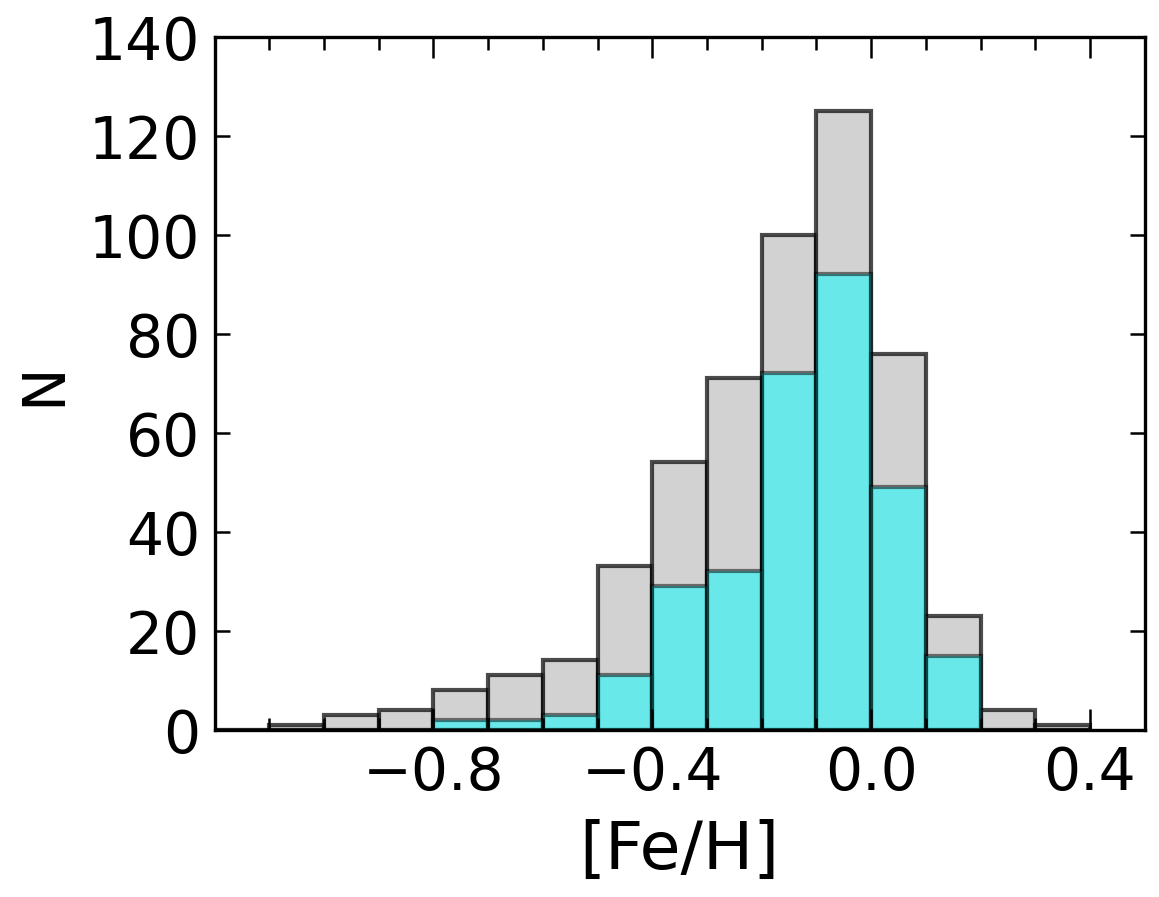} 
        \includegraphics[width=0.49\columnwidth]{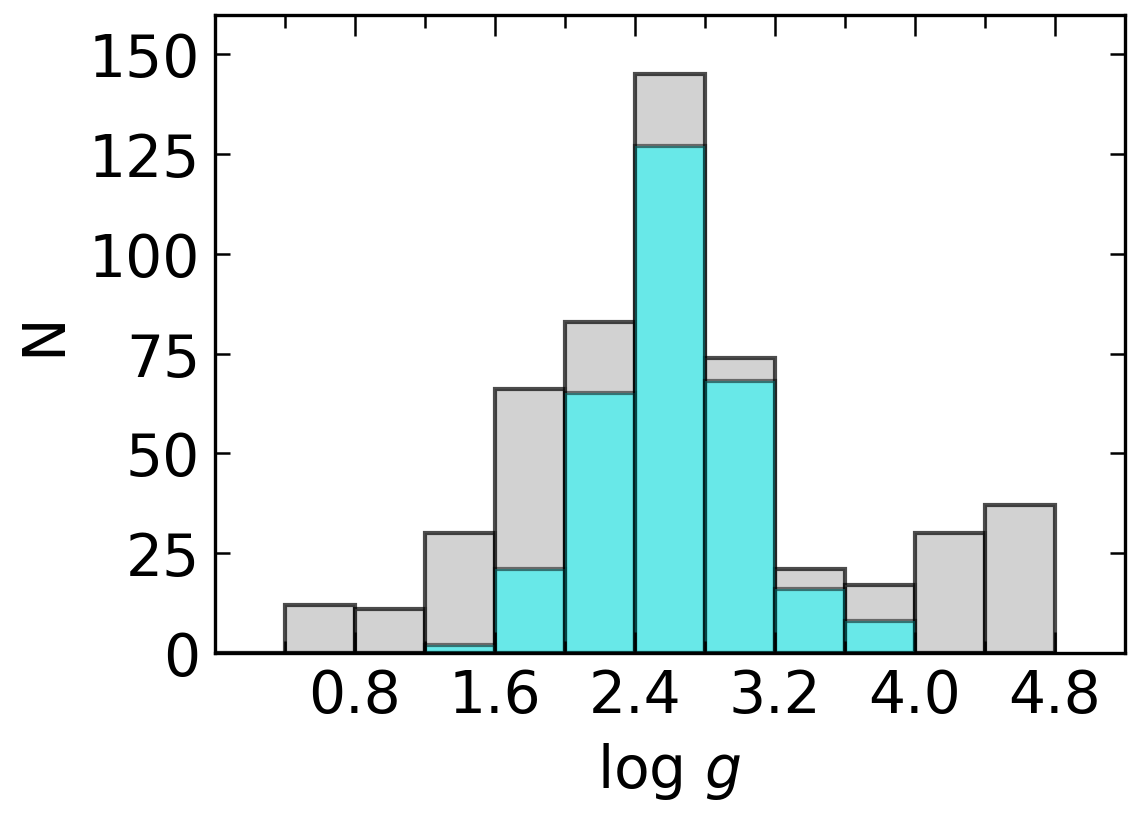}
    \includegraphics[width=0.49\columnwidth]{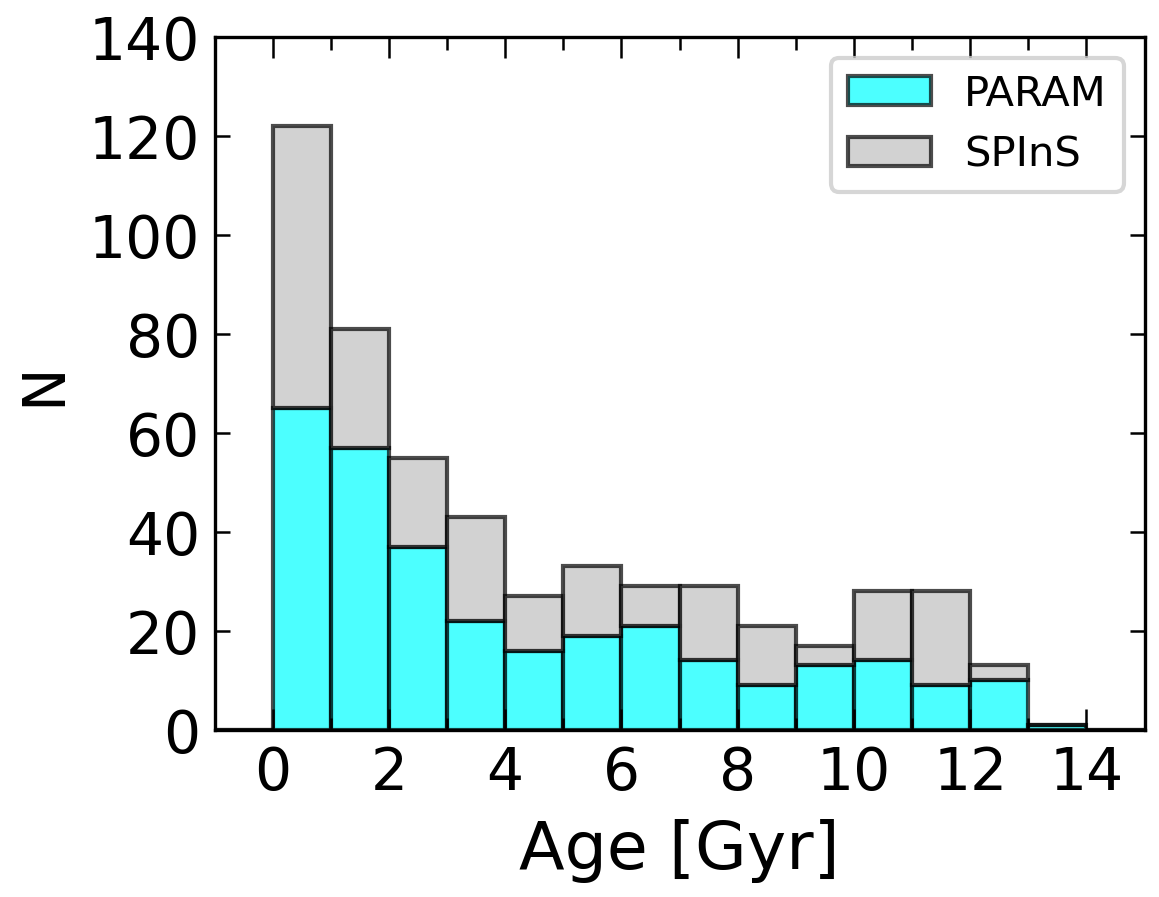} 
    \caption{Distributions of stellar parameters in the sample stars. The stars whose ages were determined with the PARAM code are shown in cyan, and those whose ages were determined with the SPInS code are shown in grey.  }
    \label{histo-all}
\end{figure}

\begin{figure}
    \centering
     \includegraphics[width=1.00\columnwidth]{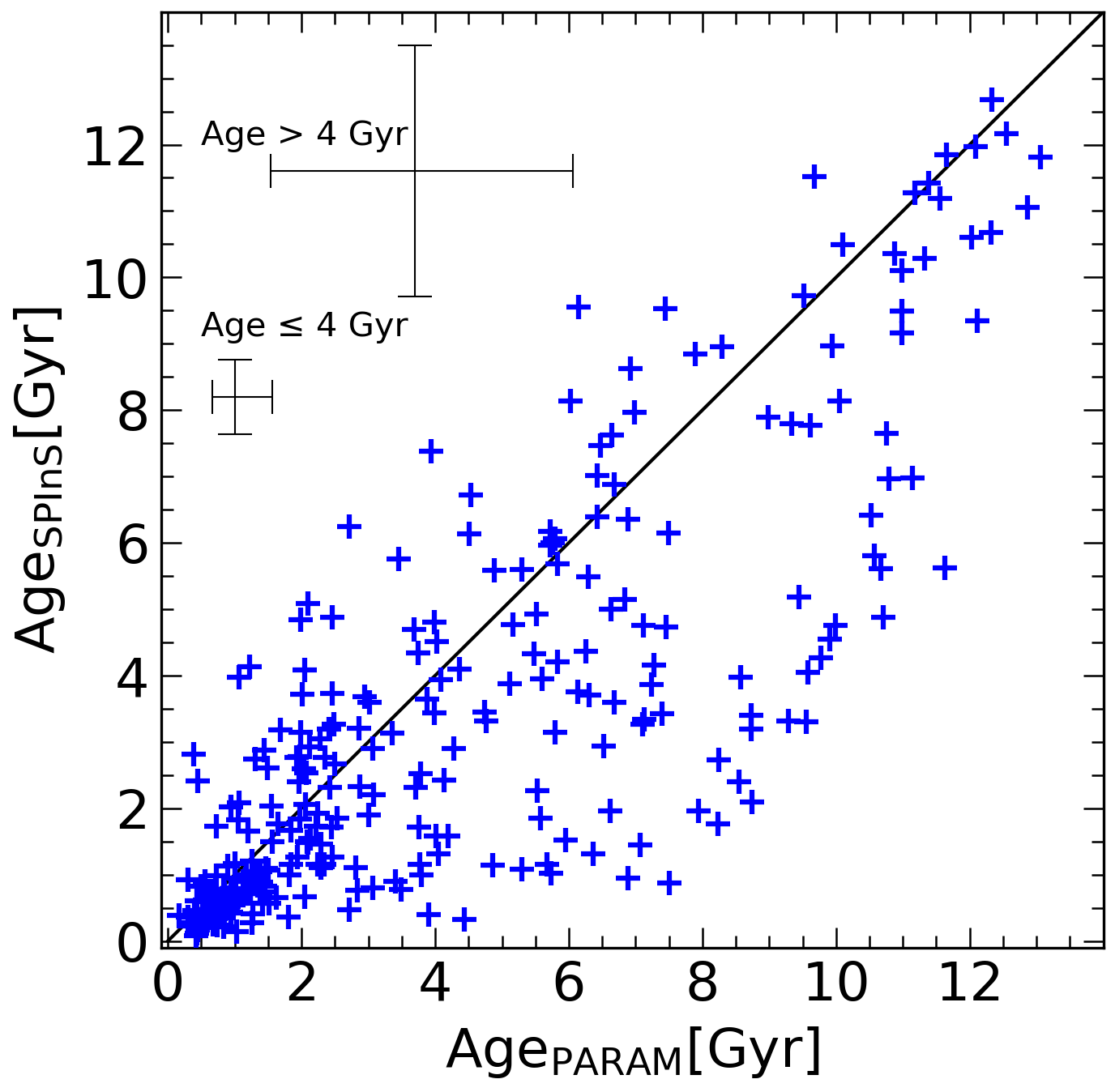}
    \caption{Comparison of ages determined with the PARAM and SPInS codes. The averaged uncertainties are indicated for stars with ages smaller and larger than 4~Gyr.}
    \label{age-compar}
\end{figure}

In this work, the fraction of stars with asteroseismic ages was higher than in \citetalias{Pakstiene2026}. The success rate was increased by the continued flow of new data and the expanded variety of cadences.  In Fig.~\ref{histo-all}, we show the distributions of stars with asteroseismic ages across the stellar atmospheric parameters. In this study, for stars without asteroseismic age determinations, we explored the possibility of using the isochronal ages.
In Fig.~\ref{age-compar}, we show a comparison of asteroseismic ages determined with PARAM and with SPInS, and one can see that they agree quite well. We therefore adopted SPInS ages for stars that did not have asteroseismic ages. The distribution of stars with the PARAM and SPInS ages is shown in the bottom-right histogram of Fig.~\ref{histo-all}.  We also checked whether inclusion of SPInS ages changes the calculated [Y/Mg] relations with age and found no significant influence (Fig.~\ref{YMg-age}).  

In the machine-readable version of Table~\ref{table:Results}, we provide the derived atmospheric parameters, stellar ages, $\nu_{\rm max}$ and $\Delta\nu$ values, NLTE [Mg/H] and [Y/H] abundances, and other quantities determined in this study for the 528 stars. For the subsequent analysis of the relation between [Y/Mg] and age, we expanded the sample to 736 stars by incorporating data from \citetalias{Pakstiene2026}.

\subsection{ Spatial relation of [Y/Mg] with age}

\begin{figure*}
    \centering
    \includegraphics[width=0.49\textwidth]{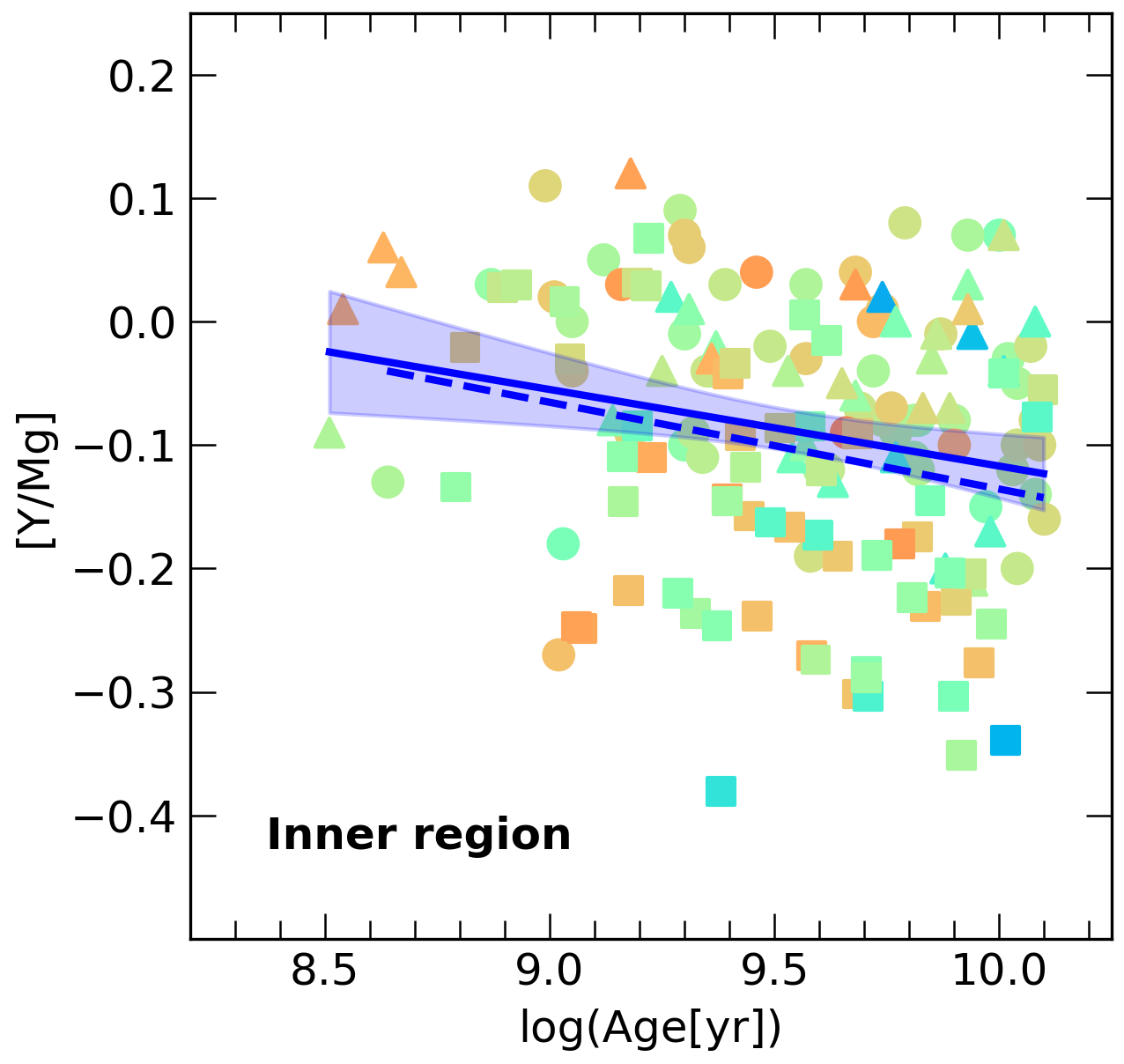}
        \includegraphics[width=0.49\textwidth]{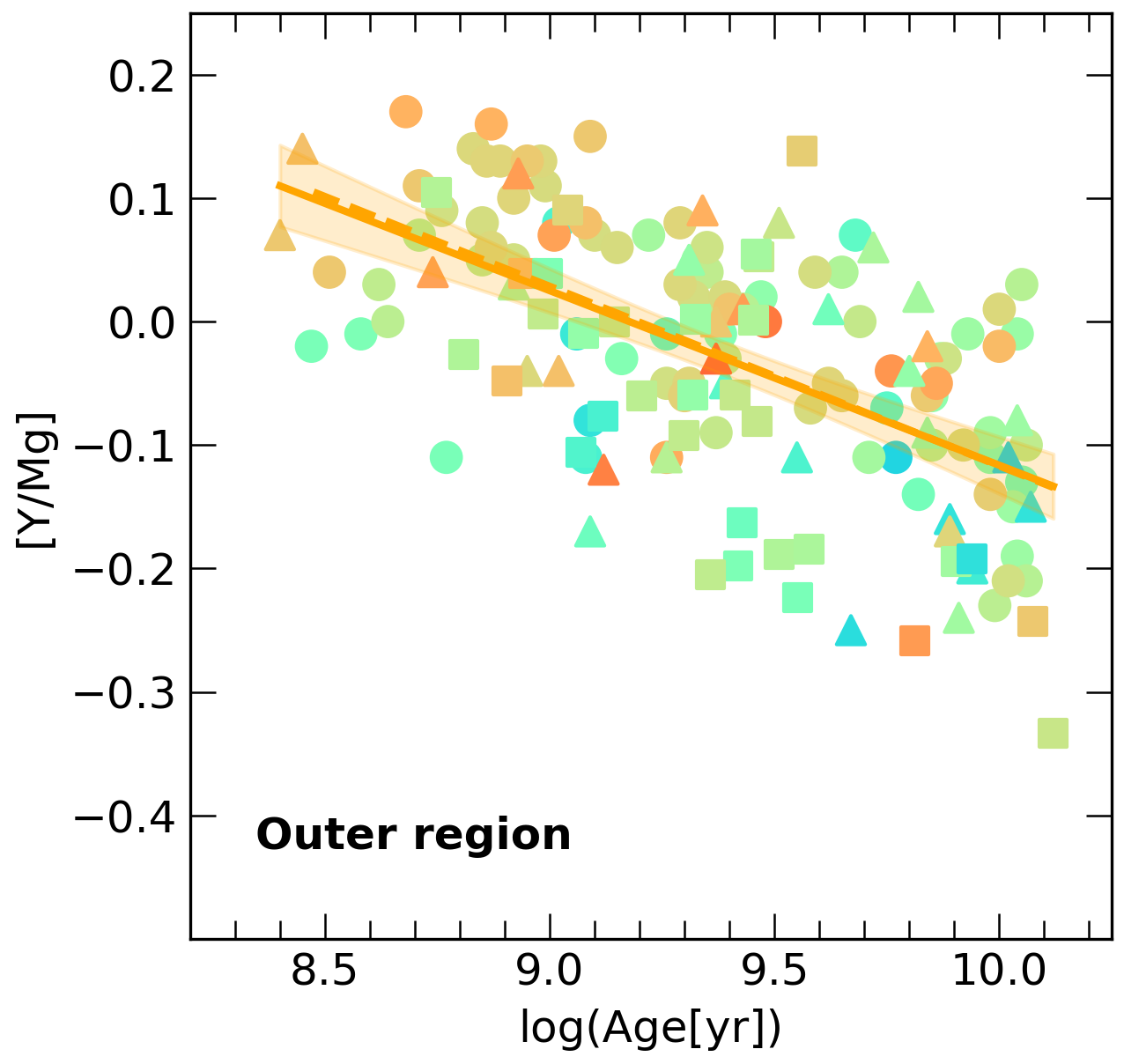}
     \includegraphics[width=0.49\textwidth]{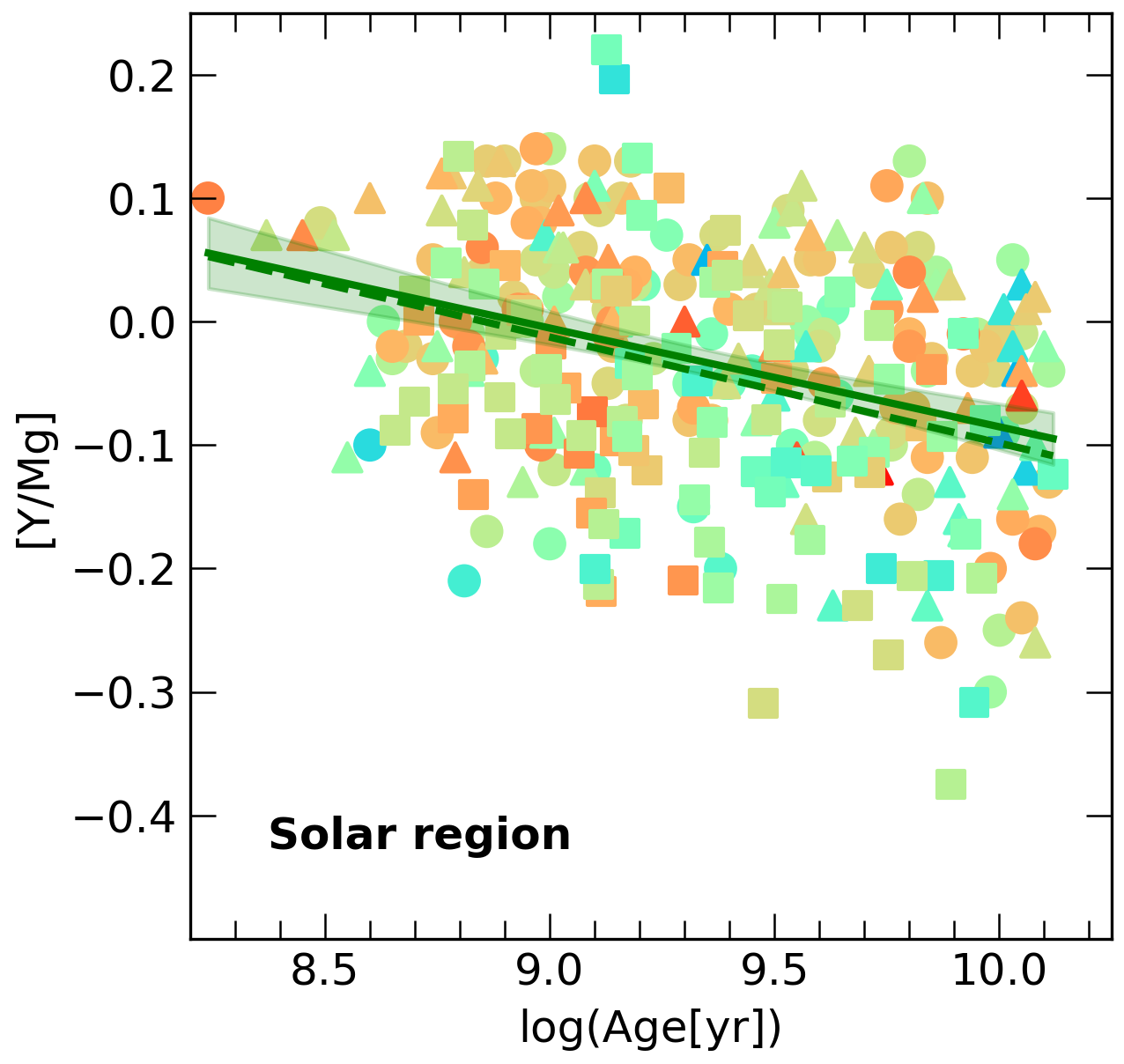}
        \includegraphics[width=0.49\textwidth]{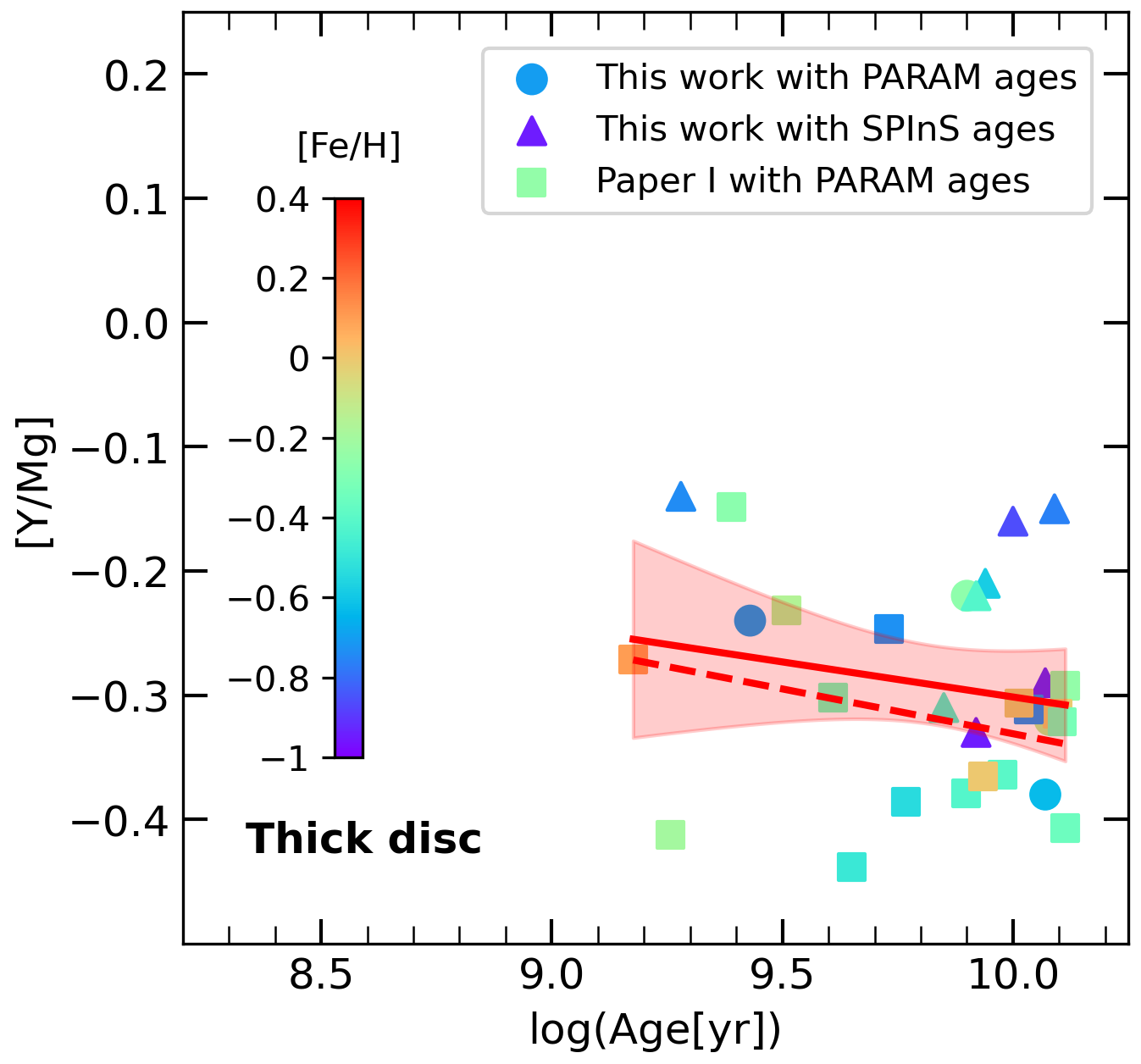}
    \caption{[Y/Mg] as a function of stellar age for thin- and thick-disc stars. Stars with $R_{\rm mean} < 7.5$~kpc are attributed to the inner Galactic disc, the solar region is at $7.5 \leq R_{\rm mean} \leq 8.5$~kpc, and the outer disc is at $R_{\rm mean} > 8.5$~kpc. Linear fits for different Galactic regions and their 95\% confidence intervals are shown. 
    The continuous line shows a fit for all the stars, while the dashed line shows the fit when stars with SPInS ages are not included. }
    \label{YMg-age}
\end{figure*}

\begin{figure*}
    \centering
    \includegraphics[width=1.0\textwidth]{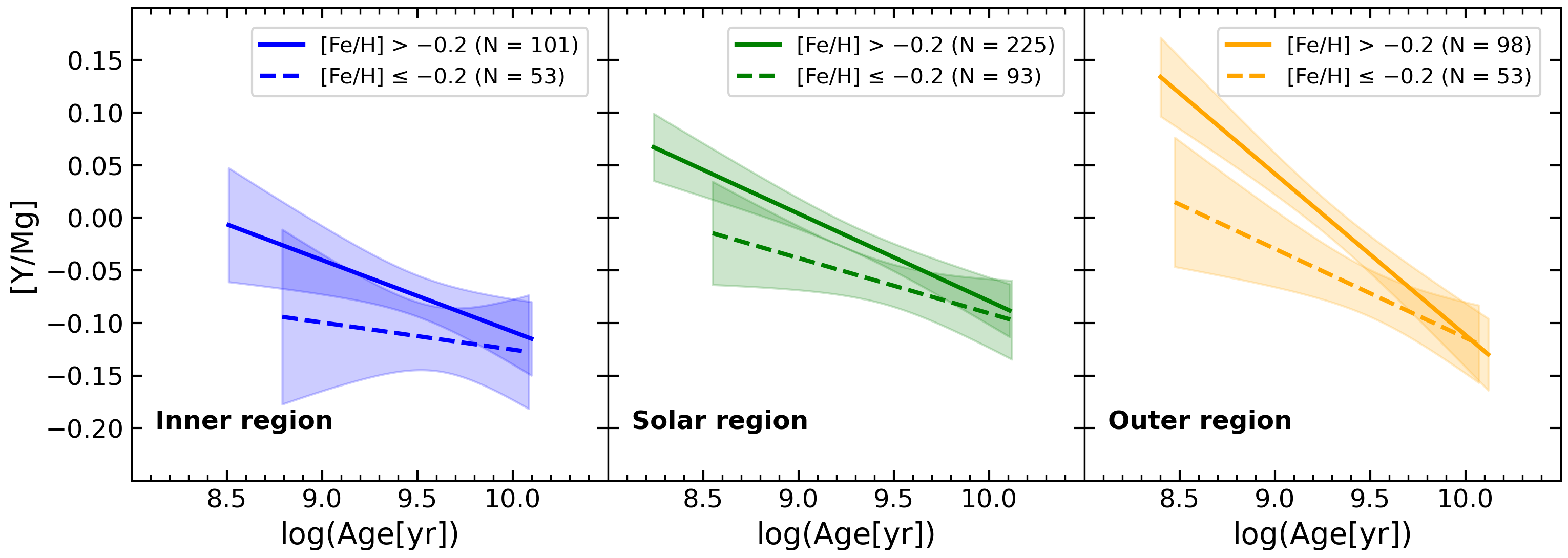}
    \caption{[Y/Mg] as a function of stellar age for the thin-disc stars divided into the inner, solar, and outer Galactic regions and divided according to metallicity. The continuous lines are for stars with [Fe/H]$> -0.2$, and the dashed lines are for ${\rm [Fe/H]}\leq -0.2$. Linear fits with their 95\% confidence intervals are shown. 
    }
    \label{YMg-metallicity}
\end{figure*}

To investigate radial dependencies in the $[\mathrm{Y}/\mathrm{Mg}]$--age relation, we divided the sample into four subsamples. As in \citetalias{Pakstiene2026} and references therein, stars with $R_{\mathrm{mean}} < 7.5$~kpc were assigned to the inner region, those with $7.5 \le R_{\mathrm{mean}} \le 8.5$~kpc define the solar region, and stars with $R_{\mathrm{mean}} > 8.5$~kpc trace the outer region. Stars younger than 0.17~Gyr were not included to calculate the relations due to statistically very small samples. Thick-disc stars were treated as a separate population based on their kinematic properties. The resulting $[\mathrm{Y}/\mathrm{Mg}]$--age relations are shown together with their linear fits in Fig.~\ref{YMg-age}. Different symbols indicate stars with asteroseismic ages determined using the PARAM code in this work and in \citetalias{Pakstiene2026} and stars with isochronal ages determined using the SPInS code. 
We checked whether the inclusion of stars with isochronal ages affects the derived relations, and no significant differences were found. 

These chemical clocks display systematic variations across the Galactic disc, reflecting differences in star formation and enrichment histories. In our sample, this behaviour is seen in the $[\mathrm{Y}/\mathrm{Mg}]$--age trend (Fig.~\ref{YMg-age}) and is represented by linear relations for distinct Galactic regions and populations (Eqs.~\ref{eq:inner_logAge_YMg} -- \ref{eq:thick_logAge_YMg}):

\begin{itemize}

    \item
Inner region (154 stars) \\
\vspace{-0.22in}
\begin{align}
{\rm \log\,Age = -16.129\,[Y/Mg]} + 8.113, {\rm PCC=-0.22}.
\label{eq:inner_logAge_YMg}
\end{align}

\item
Solar region (318 stars)\\
\vspace{-0.22in}
\begin{align}
{\rm \log\,Age = -12.5\,[Y/Mg]} + 8.938, {\rm PCC=-0.35}.
\label{eq:solar_logAge_YMg}
\end{align}

\item
Outer region (151 stars)\\
\vspace{-0.22in}
\begin{align}
{\rm \log\,Age = -7.092\,[Y/Mg]} + 9.206, {\rm PCC=-0.60}.
\label{eq:outer_logAge_YMg}
\end{align}

\item
Thick disc (29 stars)\\
\vspace{-0.22in}
\begin{align}
{\rm \log\,Age = -17.857\,[Y/Mg]} + 4.696, {\rm PCC=-0.20}.
\label{eq:thick_logAge_YMg}
\end{align}

\end{itemize}

The clear variation in slope and correlation strength between the inner, solar, and outer regions of the thin disc indicates that the sensitivity of [Y/Mg] to stellar age is not constant across the Galaxy. In particular, the steeper relation found in the outer disc (Eq. \ref{eq:outer_logAge_YMg}) suggests a stronger temporal evolution of the $s$-process contribution in regions characterised by lower star formation efficiency, while the flatter trend in the inner disc (Eq. \ref{eq:inner_logAge_YMg}) points to a more rapid chemical evolution that reduces the age sensitivity of this ratio. This behaviour is fully consistent with the interpretation presented in \citetalias{Pakstiene2026} and with previous observational studies of open clusters and field star samples \citep[e.g.][]{Casali20, Vazquez22, Ratcliffe24, Viscasillas2025}, which report similar radial dependencies of $s$-process-based chemical clocks. This supports the scenario in which the inner and outer parts of the Galactic thin disc do not follow the same
chemical evolution \citep{Snaith2015}. From the theoretical perspective, such trends are reproduced by the multi-zone chemical evolution models by \citet{Magrini2021}, which adopt yields of asymptotic giant branch stars including magnetic mixing. However, some discrepancies persist, particularly in the inner Galaxy at younger ages, when compared with predictions from the extended three-infall model \citep{Molero25,Palla2024}. Furthermore, the weak correlation observed for the thick-disc population (Eq. \ref{eq:thick_logAge_YMg}) supports a rapid formation scenario dominated by Type~II supernovae prior to the significant contribution of low-mass asymptotic giant branch stars \citep{Tautvaisiene2021}. In this framework, the transition from the nearly flat [Y/Mg] relation in the thick disc to increasingly steeper gradients in the inner, solar, and outer thin-disc regions reflects a clear radial trend in star formation efficiency. While the inner regions formed rapidly from the gas reservoir left by the thick disc \citep{Katz2021}, reaching a state of near chemical saturation, the outer thin disc continued to evolve its chemical clock at a much slower pace.

\subsection{Metallicity effects in [Y/Mg] and age relations}

The production of yttrium, as an $s$-process element, depends on the number of neutrons available per iron seed \citep{Kappeler11}. At lower metallicities, a higher number of neutrons per iron seed favours the synthesis of heavier $s$-process elements, whereas at higher metallicities, fewer neutrons are available per seed, leading to the production of lighter $s$-process elements such as yttrium. This has been confirmed by observations (e.g. \citealt{Feltzing17, Delgado19, Casali20, Vitali24}) finding different yttrium abundance in relation to age.

\begin{figure}
    \centering
    \includegraphics[width=1.0\columnwidth]{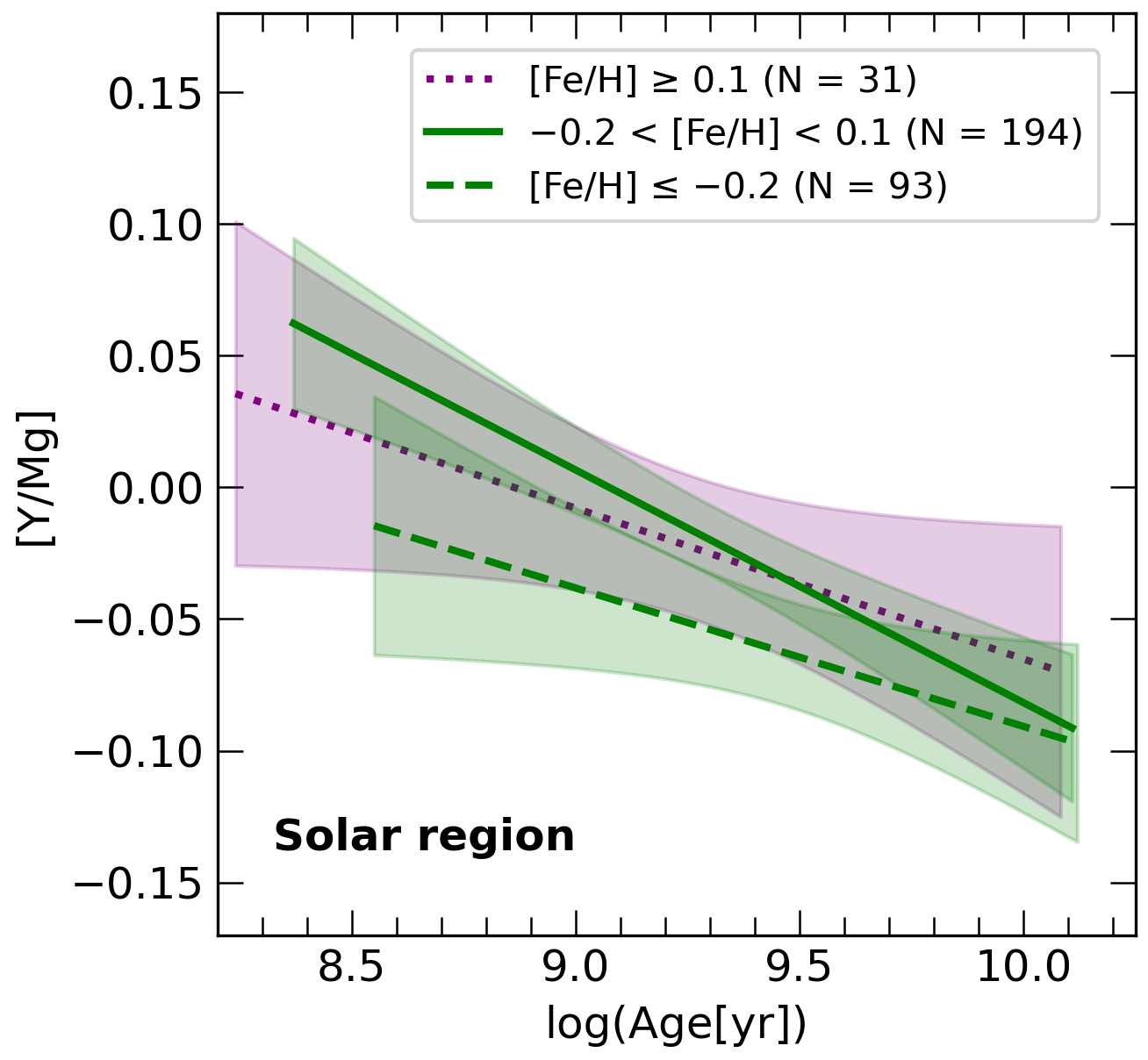}
    \caption{[Y/Mg] as a function of stellar age for the thin-disc stars in the solar region divided according to metallicity. The dotted line is for stars with [Fe/H]$>0.1$, the continuous line is for stars with $-0.2<{\rm [Fe/H]}< 0.1$, and the dashed line is for stars with ${\rm [Fe/H]}\leq -0.2$. Linear fits with their 95\% confidence intervals are shown. 
    }
    \label{YMg-metallicity-Solar}
\end{figure}

However, there is a lack of studies analysing how [Y/Mg] ratios change over time while taking into account both galactocentric distances and metallicities. For open clusters, a study by \citet{Vazquez22} computed weighted multilinear regressions. However, the sample contained quite a few clusters older than 5~Gyrs. In this context, we decided to test how the [Y/Mg] versus age relations depend on both the metallicity and the mean galactocentric distances for the field stars in a wider age interval.

In Fig.~\ref{YMg-metallicity} we depict the dependencies of [Y/Mg] versus  log(Age[yr]) as relations divided into three Galactic thin-disc regions by $R_{\mathrm{mean}}$ and into two different metallicity regimes of ${\rm[Fe/H]}> -0.2$ and ${\rm[Fe/H]}\leq -0.2$. In all three $R_{\mathrm{mean}}$ regions, we find the same tendencies -- more metal rich stars on average show larger [Y/Mg] ratios throughout the age range compared to the more metal poor stars in our sample. However, the two tendencies in each graph are not parallel. At older ages, we see that the [Y/Mg] in the relations are closer, while as the age becomes younger, the separation between the fits becomes larger. This is visible most prominently in the outer region.

Trends very similar to those found in our study were reported by \citet{Delgado19}. They, however, found that stars with ${\rm[Fe/H]}> 0.2$ have a flat [Y/Mg] and age relation, lying between stars with $-0.2<{\rm [Fe/H]}<0.2$ and those with ${\rm[Fe/H]}< -0.2$.  
\citet{Casali20} found the flattening of the [Y/Mg] relation with age already starts for stars with ${\rm[Fe/H]}> 0.1$. 
They divided stars into four metallicity regions and found that as metallicity increases, the [Y/Mg] on average becomes higher, but the highest metallicity stars show a much flatter relation, which at younger ages have lower [Y/Mg] ratios than in the more metallicity poor range.

Since we had 31 stars with ${\rm[Fe/H]}\geq 0.1$ in our sample of stars in the solar $R_{\mathrm{mean}}$ region, we compared their [Y/Mg]--age relation with those for stars with $-0.2<{\rm [Fe/H]}<0.1$ and with ${\rm[Fe/H]}\leq -0.2$. Figure~\ref{YMg-metallicity-Solar} shows the determined relations. The sample of stars with ${\rm[Fe/H]}>0.1$ is quite small, but the flattened relation is visible. In the inner and outer regions, we had just five and seven stars, respectively.    

Thus, it is clear that there is a tendency that when the metallicity increases, [Y/Mg] shows higher values throughout the age range. However, at supersolar metallicity ([Fe/H] somewhere above 0.1~dex) this tendency may not hold, and the trend could become flatter than for solar metallicity stars with lower [Y/Mg] values at young ages and higher values at old ages. The coefficients and PCC of the [Y/Mg] versus age relations obtained in this work are listed in Table~\ref{table:coeff}.  

 \begin{table}
 \caption{Coefficients and PCC for the relations ${\rm \log\,Age = m\cdot[Y/Mg] + c }$ in the three regions of the Galactic thin disc and for different metallicities.}
 \label{table:coeff}
\begin{tabular}{lccc}
\hline
 \hline 
 [Fe/H] & m & c & PCC \\
 \hline
  \noalign{\smallskip}
 \multicolumn{4}{c}{$R_{\rm mean} < 7.5$~kpc} \\
 $-0.2<{\rm [Fe/H]}<0.1$ & $-15.385$ & $8.369$ & $-0.25$  \\
 ${\rm[Fe/H]}\leq -0.2$ & $-14.706$ & $8.412$ & $-0.08$ \\
 \hline
  \noalign{\smallskip}
 \multicolumn{4}{c}{ $7.5 \leq R_{\rm mean} \leq 8.5$~kpc } \\
 ${\rm[Fe/H]}\geq 0.1$  & $-17.544$ & $8.895$ & $-0.35$ \\
 $-0.2<{\rm [Fe/H]}<0.1$ & $-11.364$ & $9.102$ & $-0.37$ \\
 ${\rm[Fe/H]}\leq -0.2$ & $-18.868$ & $8.208$ & $-0.22$ \\
 \hline
  \noalign{\smallskip}
 \multicolumn{4}{c}{$R_{\rm mean} > 8.5$~kpc }\\
 $-0.2<{\rm [Fe/H]}<0.1$ & $-6.667$ & $9.267$ & $-0.64$ \\
 ${\rm[Fe/H]}\leq -0.2$ & $-11.765$ & $8.612$ & $-0.40$ \\
\hline
\end{tabular}
\end{table}

Our study clearly shows that Galactic regions and overall metallicity must be taken into account when analysing chemical abundance–age relations.  To come closer to more robust fits, NLTE deviations must be taken into account for chemical elements, and precise ages, preferably asteroseismic, must be used. The possibility of using  methods involving neural networks remains open (e.g. \citealt{Moya2022}, \citealt{Tamames-Robero2025}). 
The work on this subject has to continue, as [Y/Mg], or any other abundance ratio, can currently only be used as an empirical age indicator rather than as a universal age-determination tool.

\section{Summary and conclusions}

In this work, we have characterised the spatial variations of the empirical [Y/Mg]--age relation across the Galactic discs. To increase the reliability of this chemical clock, we significantly enlarged our previous dataset of 208 stars (\citetalias{Pakstiene2026}) by analysing new high-resolution spectra for 528 field stars observed at the Molėtai Astronomical Observatory. As in the previous study, Mg and Y abundances were determined via spectral synthesis of multiple spectral features while accounting for NLTE effects. The timescale of our `chemical clock' was anchored using fundamental asteroseismic ages for stars showing solar-type pulsations and cross-checked isochrone-based ages for the remainder of the sample. 
Combined with our previous data, this yielded a comprehensive sample of 736 Galactic field stars with asteroseismic ages determined for 525 of them. 
Our main conclusions based on this high-precision dataset are summarised as follows:

\begin{itemize}
\item
The [Y/Mg] versus age relations display systematic variations across the Galactic discs, reflecting differences in star formation and enrichment histories.

\item 
The transition from a nearly flat relation in the thick disc to progressively steeper gradients from the inner to the outer thin disc reflects a clear radial dependence of star formation efficiency across the Galaxy.

\item 
The stronger temporal evolution observed in the outer disc suggests a larger contribution from the $s$-process in regions with lower star formation efficiency, whereas the flatter trend in the inner disc indicates a faster chemical evolution that reduces the age sensitivity of this ratio.

\item
The weak [Y/Mg]–age correlation in the thick-disc population supports a rapid formation scenario dominated by Type~II supernovae prior to the significant contribution of low-mass asymptotic giant branch stars.

\item
There is a tendency for  [Y/Mg] to increase with metallicity across the age range.
However, at supersolar metallicity (${\rm[Fe/H]}>0.1$) this tendency may not hold, and the trend could become flatter than for solar metallicity stars with lower [Y/Mg] values at young ages and higher values at old ages. 

\end{itemize}

In general, our results reinforce the conclusion that the [Y/Mg] ratio cannot be treated as a universal chemical clock. Its calibration must explicitly account for the Galactic environment, metallicity, and star formation history. Further work is required, as chemical clocks are valuable not only as age indicators but also as tools for tracing the history of the chemical enrichment of the Galaxy.

\section*{Data availability}

The complete versions of Table~\ref{table:Results} and Table~\ref{table:AtomicData} are available via anonymous FTP at the CDS.

\begin{acknowledgements}

We acknowledge funding from the Research Council of Lithuania (LMTLT, grant No. S-MIP-23-24). We thank the anonymous referee for very helpful suggestions.
This paper includes data collected by the TESS mission. Funding for the TESS mission is provided by the NASA's Science Mission Directorate. This work has used data from the European Space Agency (ESA) mission {\it Gaia} (\url{https://www.cosmos.esa.int/gaia}), processed by the {\it Gaia} Data Processing and Analysis Consortium (DPAC, \url{https://www.cosmos.esa.int/web/gaia/dpac/consortium}). Funding for the DPAC has been provided by national institutions, in particular the institutions participating in the {\it Gaia} Multilateral Agreement.

\end{acknowledgements}

\bibliographystyle{aa} % style aa.bst
\bibliography{biblio.bib} % your references Yourfile.bib

\onecolumn
\begin{appendix}

\section{Machine readable tables of results}

%%%\begin{tiny}==
 \begin{longtable}{llll}
 \caption{Parameters of the stars.}
 \label{table:Results}\\
 \hline
 \hline
 Col & Label & Units & Explanations\\
 \hline
 1      & ID                 & --          & Tycho-2 catalogue identification\\
 2      &Teff                & K    & Effective temperature\\
 3      &e\_Teff               &   K    & Uncertainty in effective temperature\\
 4       &logg                &  dex   & Surface gravity\\
 5      &e\_logg              &  dex   & Uncertainty in surface gravity\\
 6       &[Fe/H]              &  dex   & Metallicity \\
 7      &e\_[Fe/H]            &  dex   & Uncertainty in metallicity \\
 8      &Vt                  &   km s$^{-1}$  & Microturbulence velocity\\
9      &e\_Vt                &  km s$^{-1}$   & Uncertainty in microturbulence velocity\\
10     & [Mg/H]                  & dex    & NLTE magnesium abundance \\
11     & $e$\_[Mg/H]                  & dex    & Uncertainty in magnesium abundance \\
12     & [Y/H]                  & dex    & NLTE yttrium abundance \\
13     & $e$\_[Y/H]                  & dex    & Uncertainty in yttrium abundance \\
14     & [Y/Mg]                 & dex  & [Y/Mg] ratio \\
15       & Age\_PARAM          & Gyr         & Asteroseismic age determined using PARAM \\
16       & e\_lo\_Age\_PARAM        & Gyr         & Lower uncertainty (68$\%$ credible interval) in the age determination with PARAM \\
17       & e\_up\_Age\_PARAM        & Gyr         & Upper uncertainty (68$\%$ credible interval) in the age determination with PARAM \\ 
18       &  $\nu_{\rm{max}}$  &   $\mu$Hz              & Frequency at maximum power \\
19       &  $e$\_$\nu_{\rm{max}}$   &    $\mu$Hz          & Uncertainty in frequency at maximum power \\
20       &  $\Delta\nu$       &       $\mu$Hz            & Large frequency separation \\
21      &  $e$\_$\Delta\nu$       &  $\mu$Hz    & Uncertainty in large frequency separation \\
22    & Age\_SPInS              & Gyr    & Age determination using SPInS \\
23    & $e$\_Age\_SPInS         & Gyr    & Uncertainty of age deremination using SPInS \\
24     & Disc  & -- & 0 -- Thin disc, 1 -- Thick disc, 3 -- Halo, 4 -- Unclassified \\
25     & R\_mean    &  kpc  & Mean galactocentric distance \\
26     & e\_R\_mean    &  kpc  & Uncertainty in the mean galactocentric distance \\
27     & Z\_max        &  kpc  & Maximum distance from the Galactic plane \\
28     & e\_Z\_max        &  kpc  & Uncertainty in the maximum distance from the Galactic plane \\
 \hline
 \end{longtable}
 \tablefoot{Full table is available at the CDS}

\begin{longtable}{llll}
\caption{Atomic data of spectral lines used in the abundance analysis}
\label{table:AtomicData}\\
\hline
\hline
Col & Label & Units & Explanation\\
\hline
1  & Element       & --  & Chemical element\\
2  & Ion       & --  & Ionisation stage\\
3  & Wavelength    & nm  & Central wavelength of the spectral line\\
4  & E$_{\rm low}$ & eV  & Lower excitation potential of the transition\\
5  & $\log gf$     & dex & Oscillator strength of the transition\\
\hline
\end{longtable}
\tablefoot{Full table is available at the CDS.}

 \centering
%%%\end{tiny}
\end{appendix}

\end{document}